**Biodose Tools updates for criticality accidents and interlaboratory comparisons.**

Anna Francès-Abellán[1], David Endesfelder[2], Alfredo Hernández[3], Gemma Armengol[1], Joan Francesc Barquinero[1*].

[1] Unit of Biological Anthropology, Department of Animal Biology, Plant Biology and Ecology, Universitat Autònoma de Barcelona, Bellaterra E-08193, Catalonia, Spain.

[2] Department of Effects and Risks of Ionising and Non-Ionising Radiation, Federal Office for Radiation Protection (BfS), Oberschleissheim, Germany.

[3] Independent Researcher, Norwich, UK.

* Corresponding author: Joan Francesc Barquinero. Unit of Biological Anthropology, Department of Animal Biology, Plant Biology and Ecology, Universitat Autònoma de Barcelona, Bellaterra E-08193, Catalonia, Spain. email: francesc.barquinero@uab.cat

**ORCID NUMBERS**

Anna Francès-Abellán, 0000-0003-4738-1712

David Endesfelder, 0009-0008-8942-8017

Alfredo Hernández, 0000-0002-2660-4545

Joan Francesc Barquinero, 0000-0003-0084-5268

Gemma Armengol, 0000-0003-2345-1106

**BIOGRAPHICAL NOTE**

**Anna Francès-Abellán, Ph.D. student,** Unit of Biological Anthropology, Department of Animal Biology, Plant Biology and Ecology, Universitat Autònoma de Barcelona, Bellaterra, Spain

**David Endesfelder, Ph.D., Biomathematician,** Federal Office for Radiation Protection (BfS), Unit: Biological Dosimetry, Oberschleissheim, Germany.

**Alfredo Hernández, MSc, Physicist and Data Scientist,** Independent Researcher, Norwich, UK.

**Joan Francesc Barquinero, Ph.D., Biologist, Full Professor**, Unit of Biological Anthropology, Department of Animal Biology, Plant Biology and Ecology, Universitat Autònoma de Barcelona, Bellaterra, Spain.

**Gemma Armengol, Ph.D., Biologist, Associate Professor,** Unit of Biological Anthropology, Department of Animal Biology, Plant Biology and Ecology, Universitat Autònoma de Barcelona, Bellaterra, Spain.

**ABSTRACT**

**Purpose:** Since its initial release, the aim of Biodose Tools was to offer an easy-to-use platform to perform the mathematical calculations needed in biological dosimetry. This update 3.7.1, mainly focuses on new features related to large-scale emergency responses, like criticality accidents dose estimation and laboratory networks.

**Material and Methods:** Biodose Tools has been developed using the R programming language. The current version (3.7.1) uses the same external dependencies as version 3.6.1 (released November 2022) while integrating three new external packages to support the new functionalities.

**Results:** Version 3.7.1 introduces different new modules: (a) a characteristic limits module that calculates decision thresholds and detection limits following ISO19238:2023 standards, and offers statistical tests to compare rates between suspected exposure cases and control data; (b) an enhanced dose estimation module which supports multiple dose assessments for dicentric and translocation assays for various exposure scenarios—acute, protracted, and highly protracted—as well as whole and partial-body exposures; (c) a criticality accidents module for multiple dose estimations using dicentrics in mixed gamma-neutron exposure scenarios (e.g., nuclear detonations); and (d) an Interlaboratory comparison module that automates the evaluation and comparison of dose estimates across laboratories.

**Conclusions:** Biodose Tools (https://biodosetools.reneb.bfs.de/) continues to evolve in response to the dynamic needs of the biological dosimetry community, contributing to the preparedness and consistency in emergency response and routine applications.

# 1. INTRODUCTION

Among the various tools available in the field of radiological protection, biological dosimetry aims to estimate the absorbed dose after a suspected overexposure to ionizing radiation by analyzing a biological marker (IAEA 2011; Blakely et al. 2024), and plays a critical role in the medical management of potentially exposed individuals (Ainsbury et al. 2022). The estimation of a dose and the corresponding uncertainties rely on mathematical modeling and the assumption of a statistical probability distribution of specific biomarkers with a well-established dose-response relationship. Given the frequent complexity of radiation exposure scenarios, biological indicators not only aid in dose estimation but also could provide critical insights into exposure circumstances, complementing physical dose reconstruction, when necessary (Grégoire et al. 2018). Most biomarkers used in biological dosimetry are linked to DNA damage and are typically applied through cytogenetic techniques, including the analysis of dicentric chromosomes, translocations, and micronuclei. Detailed information of those assays applied in biological dosimetry can be found in IAEA technical document (IAEA 2011), and in several ISO standards (ISO19238 2023 for dicentrics, ISO20046 2019 for translocations and ISO17099 2024 for micronuclei assays).

In order to deal with the statistical procedures involved in biological dosimetry some auxiliary software has been developed since the 1980's (Ainsbury and Barquinero 2009). Notable examples include MLPOL (Szłuińska et al. 2005), MLREG (Bundesamt für Strahlenschutz 1996), DOSGEN (Garcia and Zequera 1996), CABAS (Deperas et al. 2007), and Dose Estimate (Ainsbury and Lloyd 2010). While these programs have made significant contributions to the field, they were developed on closed-source platforms, resulting in some limitations. In particular, they often employed a single methodology reducing their adaptability and, because modifications were restricted to the original developers, the community was unable to customize or extend the software to meet evolving research requirements.

On this basis, in 2023, the RENEB (Running the European Network of Biological Dosimetry and Retrospective Physical Dosimetry) association (Kulka et al. 2012) developed *biodosetools* (styled as Biodose Tools) to become an intuitive resource for performing the necessary tests and calculations required by biological dosimetry laboratories worldwide (Hernández et al. 2023). The software can be used as web version (https://biodosetools.reneb.bfs.de/) and is also publicly available as an R package (R Core Team 2024) on the Comprehensive R Archive Network (CRAN) at: https://cran.r-project.org/web/packages/biodosetools/index.html. The source code is accessible on GitHub at: https://github.com/biodosetools-team/biodosetools licensed under the GPL-3.0 (GNU General Public License v3.0). The most important goal of this project was to provide user-friendly tools with graphical user interfaces (GUIs) that facilitate obtaining harmonized statistical calculations. It is designed as a modular open-source software, allowing users to contribute with improvements, address new challenges, and expand its functionality in a collaborative environment.

In the previous version (3.6.1), Biodose Tools supported dose-effect fitting and dose estimation for the dicentric and translocation assays under several scenarios. However, since its release, new needs have promoted the proposal and development of additional modules to extend the application's capabilities and address emerging challenges in the field. The potential evil-intended use of radioactive materials or large-scale radiation incidents involving numerous exposed individuals requires the multiple management of a great number of samples, an aspect that was evaluated in the MULTIBIODOSE project (Jaworska et al. 2015; Monteiro Gil et al. 2017). Moreover, in case of a nuclear detonation the capability to estimate doses under criticality accident circumstances is needed. The criticality accident is the result of an uncontrolled and sustained chain fission reaction with a rapid release of energy, primarily in the form of heat, along with both neutron and gamma radiation (Barbry and Fouillaud 2001). Noteworthy, one of the shortcomings of the first version of Biodose Tools was that it did not have a module for estimating doses in case of criticality accidents, a situation that was already included in the IAEA technical report in 2011 (IAEA 2011), nor multiple dose estimation capacity.

Another aspect that Biodose Tools had to address was the interface and calculations needed to evaluate the results of interlaboratory comparisons (ILCs). The limited capacity of biological dosimetry laboratories to conduct simultaneous analyses has stimulated the creation of laboratory networks that can mutually assist to ensure effective cooperation. Participating laboratories must report consistency in accuracy and precision when assessing radiation doses. For this reason, ILCs are essential for standardizing methodologies and ensuring reliable dose assessments. In fact, a large number of national or international ILCs have been conducted during the last decades (Di Giorgio et al. 2011; Wilkins et al. 2015; Oestreicher et al. 2017; Endesfelder et al. 2021; Gregoire et al. 2021; Endesfelder et al. 2023; Port et al. 2023;González Mesa et al. 2024; Bucher et al. 2025). In the previous version of Biodose Tools, no module was included to facilitate and harmonize the calculations required in an ILC.

Additionally, Biodose Tools needed a module to calculate the characteristic limits - decision threshold and detection limit - to make the best and informed decisions according to the tolerated false positive and negative rates, a requirement of the ISO19238 2023 standard for dicentric chromosome assay, the most frequently applied method.

The aim of this paper is to present the new updates implemented in Biodose Tools (biodosetools 3.7.1) that include: (a) statistical tests for comparing rates between the evaluated case (suspected of exposure to ionizing radiation) and the control data; (b) a characteristic limits module that computes decision thresholds and detection limits following ISO19238 2023 standards; (c) an enhanced dose estimation module for dicentric and translocation assays, which supports multiple dose assessments for acute, protracted and highly protracted scenarios, and both whole and partial-body exposures; (d) a criticality accidents module for multiple dose estimation - using dicentrics - in scenarios involving combined gamma and neutron radiation exposure such as nuclear detonations; (e) an ILC module that automates and facilitates the evaluation of dose estimation results across laboratories; and (f) the identification and resolution of bugs from the previous version.

## 2. MATERIALS AND METHODS

### 2.1 Statistical considerations

*2.1.1 Characteristic limits*

Following ISO19238 2023 guidelines, the decision threshold and the detection limit must be calculated and reported. These measures give the lower limits of the assays for a given type I (false positives) or type II error (false negatives) respectively. In simple terms, if the observed number of dicentrics is higher than the decision threshold, there is evidence for a significant difference between the unirradiated background data and the observed numbers of dicentrics. Furthermore, if the observed number of dicentrics is higher than the detection limit, the probability to infer false negative conclusions, or to assume that a person has not been exposed when it has, is lower than the chosen type II error rate $\beta$ (see ISO19238 2023 for a more detailed explanation).

The decision threshold is calculated by finding the first number of dicentrics ($y^*$) which the R function *poisson.test()* shows a significant P-value ($P < 0.05$). The decision threshold is defined by $y_d = y^* - 1$. The detection limit $y_z$ is then calculated based on Eq. 1:

$$y_z = \frac{\chi^2_{2(y_d+1),(1-\beta)}}{2} \qquad (1)$$

where $\chi^2$ is the chi-squared quantile (the inverse of the chi-squared distribution); $\beta$ is the chosen type II (false negative) error; and $2(y_d+1)$ are the respective degrees of freedom. $\chi^2$ can be found in the corresponding tables or calculated with statistical software (e.g. R).

*2.1.2 Criticality accidents*

In a criticality accident, the human body is exposed to both neutron and gamma radiation. When the ratio of neutron to gamma-ray doses is known, typically through physical measurements, it is possible to estimate the individual neutron and gamma-ray doses using an iterative (IAEA 2011) or an analytical (Fornalski 2014; Słonecka et al. 2018) approach. However, the ratio can be difficult to obtain in a real-case scenario. In such cases, it can be approximated through physical dosimetry and computational modeling, or alternatively, using a Bayesian approach (Brame and Groer 2003; Słonecka et al. 2018; Słonecka et al. 2019) that infers dose values based on a prior distribution of the ratio.

It is assumed that the distribution of dicentric chromosomes in a given sample follows a Poisson distribution and that the observed chromosomal aberrations result from both neutron and gamma radiation exposure in an additive manner. Consequently, the dose-response relationship for a mixed neutron-gamma radiation field can be mathematically described as in Eq 2:

$$\lambda_{tot} = \frac{u}{w} = \lambda_{n+g}(D_n, D_g) = C + \alpha\, D_n + \beta\, D_g + \gamma\, D_g^2 \qquad (2)$$

where $\lambda_{tot}$ is the frequency of chromosome aberrations after irradiation by a mixed neutron and gamma radiation field; *u* the number of chromosome aberrations; *w* the number of cells scored in the sample; $D_n$ the estimation of the neutron dose and $D_g$ the estimation of the gamma dose.

In Biodose Tools, the estimation of $D_g$, $D_n$ and total dose ($D_{tot}$) is achieved through the analytical method (Fornalski 2014; Słonecka et al. 2018) (Eq 3-6).

$$D_g = \frac{\sqrt{\left(\alpha\frac{1-\theta}{\theta} + \beta\right)^2 + 4\gamma(\lambda_{tot} - C)} - \left(\alpha\frac{1-\theta}{\theta} + \beta\right)}{2\gamma} \quad (3)$$

$$D_n = \frac{1-\theta}{\theta} D_g \quad (4)$$

where θ is defined as in Eq 5 and *ρ* is the neutron to gamma-ray ratio:

$$\theta = \frac{D_g}{D_g + D_n} = \frac{D_g}{D_{tot}} = \frac{1}{\rho + 1} \quad (5)$$

Ultimately, the uncertainties associated with the estimated doses are quantified using the delta method as in Eq 6.

$$\sigma_{D_x} = \sqrt{\left(\frac{\partial D_x}{\partial \alpha}\right)^2 \sigma_\alpha^2 + \left(\frac{\partial D_x}{\partial \beta}\right)^2 \sigma_\beta^2 + \left(\frac{\partial D_x}{\partial \gamma}\right)^2 \sigma_\gamma^2 + \left(\frac{\partial D_x}{\partial \lambda_{tot}}\right)^2 \sigma_{\lambda_{tot}}^2 + \left(\frac{\partial D_x}{\partial C}\right)^2 \sigma_C^2} \quad (6)$$

where x={g,n} and σ is the standard deviation of each parameter. This method returns accurate results while maintaining relatively low computational demands.

*2.1.3 Interlaboratory comparison*

In this module, the Z-Score approach is used to assess the performance of each participating laboratory by assuming normality of the score and then establishing thresholds to define this performance, indicated as satisfactory, questionable or unsatisfactory. It provides a way to compare individual data points by standardizing the values allowing for meaningful comparisons.

To do so, the deviation of each laboratory's reported frequency from the robust average or the deviation of each laboratory's estimated dose from the physical dose is calculated as in Eq 7:

$$z = \frac{x_i - x_{ref}}{\sqrt{(s^*)^2 + u_{ref}^2}} \quad (7)$$

where $x_i$ is the estimated dose or the frequency of observed aberrations; $x_{ref}$ is the reference dose value for doses or the robust average ($x^*$) for frequencies; $s^*$ is the robust standard deviation and $u_{ref}$ is the standard uncertainty of $x_{ref}$. When frequencies (e.g. frequency of observed dicentrics) are evaluated, $u_{ref}$ is (Eq 8):

$$u_{ref} = \frac{1.25 s^*}{\sqrt{n}} \quad (8)$$

where n is the number of participants. When dose reference values are applied, $u_{ref}$ is considered negligible. More detailed information on the mathematical definitions and calculations of the Z-Score in biological dosimetry can be found in Di Giorgio et al. 2011.

The present update of Biodose Tools incorporates three options to obtain the robust standard deviation ($s^*$) and the robust average for frequency ($x^*$): (a) Algorithm A which applies the *hubers()* function from the MASS R-project package (Venables and Ripley 2002); (b) Algorithm B (Rousseeuw and Verboven 2002; Szewczak and Bondarzewski 2016), also named as the logistic M-estimator in section D.1.4.2 of the ISO13528 2022; and (c) Q/Hampel algorithm, described in detail in the section C5 of the ISO/IEC13528 2015 and in the German DIN 38402-45. See González Mesa et al. 2023 for more detail.

The ILC module also incorporates automatically the calculation of the deviation from the reference dose (*dev*) for *Dose* type input. The deviation is defined as the difference between the estimated dose $x_i$, and the reference value $x_{ref}$ (Eq. 9):

$$dev = x_i - x_{ref} \qquad (9)$$

To evaluate the ILC module, three laboratories (referred here as A1, A2, A3) were randomly selected from participants in the RENEB ILC 2023 (Bucher et al. 2025). For each laboratory, the corresponding calibration curves and dose estimations (three samples referred here as 1, 2, 3) were recalculated using Biodose Tools 3.7.1.

**2.2 Package implementation**

This version (3.7.1) relies on the same external packages as the previous version (3.6.1 – November 2022). This includes *dplyr* (Wickham et al. 2022), tidyr (Wickham and Girlich 2022), *rlang* (Henry and Wickham 2022), and *magrittr* (Bache and Wickham 2022). All of these packages are part of the *tidyverse* meta-package (Wickham et al. 2019). In Biodose Tools 3.7.1, *ggplot2* (Wickham 2016) is used for visualizations as well, but in some cases, the graphics package (R Core Team 2024) is used instead. Most statistical calculations are implemented with *stats* package (R Core Team 2024), otherwise, *maxLik* (Henningsen and Toomet 2011), *mixtools* (Benaglia et al. 2010), *MASS* (Venables and Ripley 2002), and *msm* (Jackson 2011) are applied. The Biodose Tools user interface is written in R *shiny* (Chang et al. 2025) using *Bootstrap 3*, via the *shinydashboard* (Chang and Borges Ribeiro 2021), *shinyWidgets* (Perrier et al. 2022), and *bsplus* (Lyttle 2021) packages, and the *golem* (Fay et al. 2022) framework. The shiny app allows to generate and download reports, which are rendered using *rmarkdown* (Allaire et al. 2014).

Three external packages have been integrated into the new modules. The *pdftools* package (Ooms, 2025) is used for text extraction, rendering, and PDF conversion for the ILC module report. The *openxlsx* package (Schauberger and Walker 2025) simplifies Excel file creation, and *readr* (Wickham et al. 2024) assists the reading of rectangular text data.

## 3. RESULTS

### 3.1 Characteristic limits

*3.1.1. Comparison case vs. control*

The first step in biological dosimetry is usually to check if there is evidence that a person has been exposed or not. This is usually done by comparing the aberration yield of a potentially exposed person to data from an unexposed control population. To provide the end-user with a simple and standardized way to perform this test, this option has been implemented in the new version as part of the characteristic limits module. When choosing *Compare case vs. control* in the method tab, a P-value and summary statistics are calculated based on the null-hypothesis that the counts of the case (suspected of exposure to ionizing radiation) and the control data are equal. A two-sided Poisson test is used (R Core Team 2024). A significant P-value suggests that the case was exposed to a dose >0 Gy and the estimation of the dose shall be performed if an appropriate calibration curve is available.

*3.1.2. Decision threshold and detection limit*

The calculation of these two parameters is available in the characteristic limits module choosing the corresponding method in the dropdown. The calculations are performed based on the assumption that the counts are Poisson distributed. The user can choose the type I error rate α (false positive rate) and the type II error rate β (false negative rate).

If possible, the user should choose the *Number of dicentrics and cells* option and provide the number of dicentrics and cells of the control data. This accounts for the uncertainty of the control data as well as for the uncertainty of the case data. If only the mean number of dicentrics per cell is available, the user should choose the option *Dicentrics per cell*. This does not account for the uncertainty of the control data and should not be used for relatively low cell numbers.

As a result, the Biodose Tools software returns a table with the decision threshold, the minimum number of significant dicentrics (y*) and the detection limit for multiple scenarios with different numbers of analyzed cells. If a calibration curve is provided, either manually or with an .rds file (R specific file type generated by Biodose Tools), the minimum resolvable dose and the dose at detection limit will be also displayed (Table 1). The table can be saved as Excel (.csv) or LaTeX (.tex) files.

**3.2 Multiple doses**

Since the new modules apply for situations where more than one individual is usually affected, the original code for dose estimation in dicentric and translocation modules has been modified to support dose assessments with multiple case data inputs. This option helps to significantly shorten processing time and optimize the number of steps the user needs to perform. The results are saved in a standardized format so they can be handled by the new ILC module. For this purpose, an option was added to include an ID for each case which then allows the matching of dose estimates from all participating laboratories.

**3.3 Criticality accidents**

*3.3.1 Curve fitting and data input options*

To perform dose estimation for criticality accidents, two curves need to be provided (a) a gamma-ray curve and (b) a neutron curve. These curves can be uploaded as .rds files if obtained with the fitting module in Biodose Tools or manually, entering the coefficients and standard errors.

The ratio of gamma ($D_g$) to neutron ($D_n$) doses has to be provided by the user. This should be obtained with the help of physicists by considering the type of accident (e.g., a nuclear detonation) and the distance and possible shielding structures between the source and the position of the victim. The case data, can be entered as (a) a table with the distribution of dicentrics (e.g., C0, C1, C2, C3, C4, ...) or (b) only the total number of cells and dicentrics. For the dicentrics distribution table (a), the multiple dose assessment feature is available.

*3.3.2 Final outputs*

Different tables for gamma, neutron and total dose estimations will be displayed in the Dose estimation box once the calculations have been performed (Figure 1). For each estimation, the associated yield and the upper and lower 95% confidence intervals (CIs) are reported along with their corresponding plot (Figure 2). A report can be generated in .pdf or .docx format containing all inputs, estimation results and graphical resources. Results can also be saved as .xlsx or .rds files.

**3.4 Interlaboratory comparison**

*3.4.1. Data Input and Laboratory Identification*

The first step for data input in the ILC module is to specify the number of participating laboratories (Figure 3). Once this information is available, the system generates corresponding entry fields for uploading .rds files and for the respective laboratory name or code. In order to maintain the data homogeneity achieved with the standardized scoring sheet used in past RENEB's ILCs, participating laboratories should estimate all doses with the biodosetools dose estimation module and use the .rds output in the ILC module. For a successful downstream evaluation of the results, it is essential that all participating laboratories use the correct sample IDs during the process. Moreover, an option has been implemented to input irradiation conditions when the calibration curve is entered manually in the Dose estimation modules in order to retain this information for the plots. Additional fields have been incorporated to offer more complete graphical outcomes. Due to these changes, the calibration curves obtained with the first version of Biodose Tools are no longer valid so new curves must be obtained with the Fitting module.

*3.4.2. Summary Tables*

After data input, a summary table including descriptive data within the .rds files (S.I. Table 1) along with a curve table with all the information regarding the calibration curves used for dose estimation (S.I. Table 2) by the participants of the ILC, are generated automatically. Those tables can be exported in Excel (.xlsx or .csv) and LaTeX (.tex), allowing further analysis and documentation. Two additional tables with only dose and frequency estimations are also created for quick consultation.

*3.4.3. Final Outputs*

The Z-Score data input box allows the selection of the input type: either Dose (Gy) or Frequency (aberrations/cell), and the algorithm: algA, algB or QHampel. Reference doses for each sample must be provided when working with doses (Figure 4). The designated sample names that appear in the .rds file will be used in all visual representations.

The generated table (Table 2) with the results can be saved in Excel (.xlsx or .csv) and LaTeX (.tex) formats for further review.

The performance classification follows standard statistical thresholds:

- Satisfactory (green): |Z-Score| < 2

- Questionable (yellow): $2 \leq$ |Z-Score| $\leq 3$

- Unsatisfactory (red): |Z-Score| > 3

Depending on the provided data, various graphical outputs can be generated: (a) Dose estimation (Figure 5a) and Yield (i.e. frequency) (Figure 5b) plots, (b) Dispersion Index Plot (Figure 5c), displaying the variance-to-mean ratio for each dataset, (c) U-Test Plot (Figure 5d), (d) Manual and Automatic Bar Plots (Figure 6a), illustrating the distribution of different irradiation conditions across laboratories, including radiation quality (e.g., $^{60}$Co, X-rays), calibration method (e.g., air kerma) and irradiation medium (e.g., air or water), and (e) Manual and Automatic Curves Plots (Figure 6b), with the dose-response calibration curves used by each laboratory, distinguishing between manual and automatically generated curves

Additionally, if the input type Dose is selected the following plots will be generated: (f) Dose Estimation Plot for each sample (Figure 7a, d), indicating the estimated dose distribution across laboratories in a tricolor visualization: Low exposures (<1 Gy) appear in green, medium exposures (1-2 Gy) in yellow, and high exposures (>2 Gy) in red; (g) Z-Score plot with Status categories by sample (Figure 7b, e): questionable output would appear in yellow, satisfactory in green and unsatisfactory in red; (h) Deviation from reference plot with Status categories by sample (Figure 7c, f): 0.5 Gy < abs(*dev*) < 1 Gy in yellow; abs(*dev*) < 0.5 Gy in green; and abs(*dev*) > 1 Gy in red; and (i) Z-Score and deviation from reference plots for all samples together (Figure 8a, b).

If the input type Frequency is selected, the plots that will appear are: (f) Z-Score plot with Status categories by sample and (g) Z-Score plot for all samples together.

A report can be generated in .pdf format containing the summary and the Z-Score tables. Additionally, the report will contain any comment specified in the corresponding box along with all graphical resources.

**4. DISCUSSION**

Biodose Tools was developed as an application to harmonize statistical calculations related to biological dosimetry. Although it was developed within the framework of the RENEB association, the developers' intention has always been to maintain its collaborative nature beyond its network and expand the capabilities of the app. When version 3.6.1 was released, several points were considered as possible improvements to be addressed in the future. In the latest update, version 3.7.1—the largest since the initial launch of the Biodose Tools software—the most important key issues have been successfully addressed.

ILCs are regularly performed in biological dosimetry and especially within the RENEB network to maintain and improve high quality standards of biological dosimetry laboratories. The statistical evaluation of ILCs has not been standardized so far - not even in ISO21243 2022, which includes network performance in large scale accident situations - and requires certain statistical and programming skills. Recent RENEB ILCs have demonstrated that it is important to view the results of an ILC from various perspectives (Gregoire et al. 2021; Endesfelder et al. 2023; Bucher et al. 2025) to further improve the process of providing high quality results for biological dosimetry laboratories. Therefore, one of the most relevant points for an improvement of the Biodose Tools software was the design of a specific module to standardize the evaluation of those results and to provide the most used statistical comparisons as tables and plots. This novel module enables the evaluation of ILCs in biological dosimetry without deep statistical or programming skills and will help to make the evaluation faster and more standardized in future. The newly implemented option to perform dose estimates for multiple cases simultaneously will further simplify dose estimations for the participating laboratories. However, it will be required to follow exact standardized workflows to obtain dose estimates in an ILC and the best way to achieve that is to use Biodose Tools in all the steps of the process.

As indicated before, during the last decade a large number of ILCs have been performed and although the final goal is to assess the dose with the corresponding uncertainties, some ILCs have focused on the criteria to score dicentrics (Romm et al. 2014) or have evaluated both frequencies of dicentrics and estimated doses (Di Giorgio et al. 2011; Oestreicher et al. 2017; Gregoire et al. 2021). For this reason, the new ILC module allows comparing doses or frequencies. This is especially useful for new participating laboratories or those lacking a calibration curve. In addition, although in most cases Z-Scores have been calculated using Algorithm A, as the primary robust method applied in biological dosimetry ILCs (Di Giorgio et al. 2011), under certain circumstances it can have advantages to calculate Z-Scores using other algorithms (González Mesa et al. 2023) available in this Biodose Tools update. In the ISO13258 standard Algorithm B is recommended for very small sample sizes, i.e. ILCs involving a low number of participating laboratories. This algorithm should be considered as a candidate to estimate the standard deviation in ILCs when the proportion of identical values is less than half of the results. The Q/Hampel algorithm seems more appropriate to be applied when the ILC includes non-irradiated samples, or samples irradiated at low doses. Without the presence of atypical values, and for irradiated samples at medium or high doses the Algorithm A is the closest to the arithmetic Z-Score formula. In a recent ILC involving experienced and unexperienced laboratories, the Q/Hampel algorithm has been used (González Mesa et al. 2024).

The assessment of doses and corresponding uncertainties for criticality accidents, i.e. mixtures of neutrons and gamma-rays, becomes more and more important due to the increasing energy demand that leads to the expansion of nuclear power infrastructure, thereby elevating the potential for criticality accidents or even the deliberate use of nuclear weapons in the current international context. Therefore, a new module was included in Biodose Tools v3.7.1 which calculates the neutron, gamma-ray and total doses and the corresponding 95% confidence intervals for dicentric chromosomes in the case of a criticality accident. For successful dose estimations, calibration curves for neutrons and gamma-rays are required as well as a ratio of the neutron to gamma-rays. In the current version, this ratio needs to be known exactly; however, in future versions, calculations based on Bayesian statistics (Brame and Groer 2003; Słonecka et al. 2018; Słonecka et al. 2019) will be included, which allows the specification of a prior distribution and therefore uncertainty on this parameter can be included. As in the standard dose estimation module, the criticality accident module can estimate multiple cases simultaneously, providing resources for situations where big scale studies are needed.

Furthermore, the characteristic limits module has been designed to compute decision thresholds and detection limits following ISO19238 2023 standards and moreover, provide statistical tests for comparing rates between case and control data. Even this was not set as an objective at first, it became clear that there was a need for a module that could provide these parameters since approximately between 60 and 80% of potentially exposed individuals referred for analysis turns out to had received a dose below the minimum detectable level of 0.1-0.2 Gy for the dicentric chromosome assay with low Linear Energy Transfer (LET) radiation (Lloyd et al. 2000; IAEA 2011; Sun et al. 2016; UK Health Security Agency 2024). Calculation of characteristic limits is important to assess the lower limits of a biological dosimetry assay in order to decide (1) if a physical effect quantified by the measurand is present, allowing to determine if an individual has been exposed to radiation without an available calibration curve uploaded in the data input, and (2) at which confidence level the effect can be detected to make the best and informed decisions according with the tolerated false positive and negative rates. Additionally, if a curve is available, Biodose Tools will report the minimum resolvable dose and the dose at detection limit. The new module provides a user-friendly and standardized way to derive the lower limits without deep understanding of the statistics and without having programming skills, enabling laboratories performing biological dosimetry to estimate these parameters as recommended in ISO19238 2023.

For now, there are plans to integrate Bayesian dose estimation methods (Higueras et al. 2016) as well as to integrate existing software for γ-H2AX dose estimation (Einbeck et al. 2018) into Biodose Tools. Moreover, methods for the comparison of calibration curves between multiple donors are in early stages of development. Another aspect to consider is including the information provided by retrospective physical dosimetry, both to compare the results of biological dosimetry and to integrate it into the Bayesian approach.

Overall, Biodose Tools is in a state of continuous expansion, evolving to address the dynamic challenges and requirements of the field. The intention of the team has always been to serve as a platform to improve the standardization of mathematical calculations, such as the assessment of dose and uncertainties, required in biological dosimetry (Hernández et al. 2023). This project represents a joint international effort between biological dosimetry specialists such as mathematicians and biologists from different laboratories worldwide, collecting the most commonly used statistical methods to be used in each step of the fitting and dose estimation processes in a robust and user-friendly way.

## 5. CONCLUSION

The development and continuous improvement of the Biodose Tools has made it a valuable instrument in the field of biological dosimetry, particularly with regard to the increasing networking of laboratories. An important aspect of this is the harmonization and standardization of statistical analysis, which ensures the comparability of results from different laboratories.

In the updated version of Biodose Tools three new modules have been included: (a) characteristic limits, consisting of decision thresholds, detection limits and statistical tests for comparing rates between the evaluated case (suspected of exposure to ionizing radiation) and control data, (b) criticality accidents and (c) ILC. Other minor updates: (d) multiple dose estimation, (e) .xlsx save option, (f) irradiation conditions for curve manual entry and (g) several bug fixes. There are some new modules under development including: (a) Bayesian estimation, (b) γ-H2AX dose estimation and (c) comparison of different calibration data.

The application can be found in GitHub: https://github.com/biodosetools-team/biodosetools; as a web in https://biodosetools.reneb.bfs.de/ and is also publicly available as an R package (R Core Team 2024) on the Comprehensive R Archive Network (CRAN) at: https://cran.r-project.org/web/packages/biodosetools/index.html. All the information can be found on: https://www.reneb.net/.

**ACKNOWLEDGMENTS**

We would like to thank all members of RENEB association who helped in the evaluation of Biodose Tools software and provide critical insight for the improvement of the manuscript.

**FUNDING DETAILS**

This work was supported by the Consejo de Seguridad Nuclear Español under Grant PR-102-2023.

**DISCLOSURE STATEMENT**

The authors report there are no competing interests to declare.

**FIGURES**

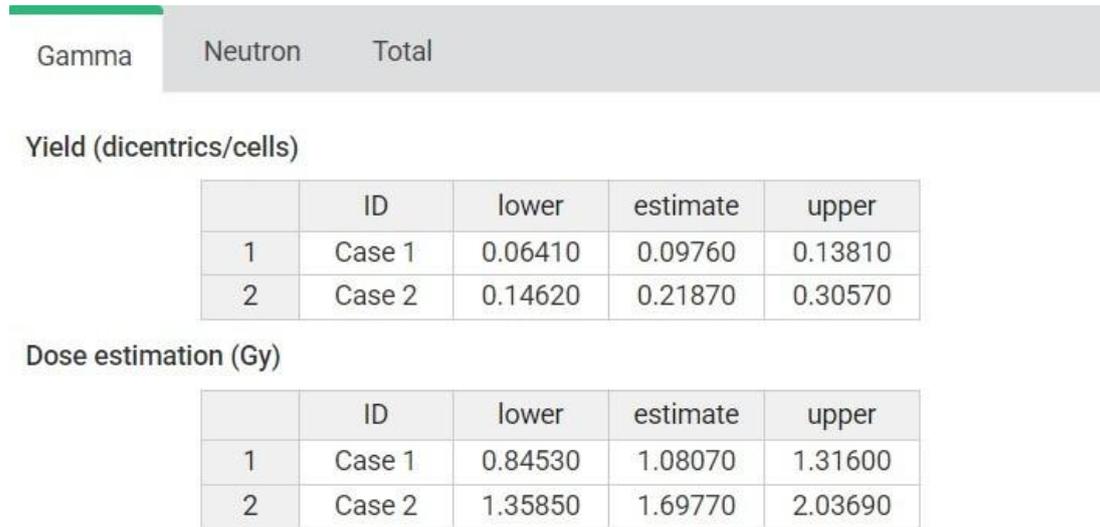

**Figure 1**. **Dose estimation results for two cases in the criticality accidents module.** The gamma, neutron and total tabs display the estimated gamma, neutron or total doses with the corresponding upper and lower 95% confidence intervals and the associated yields, respectively. Random data from Ekendahl et al. 2025 has been used in this example with a ratio of 0.72.

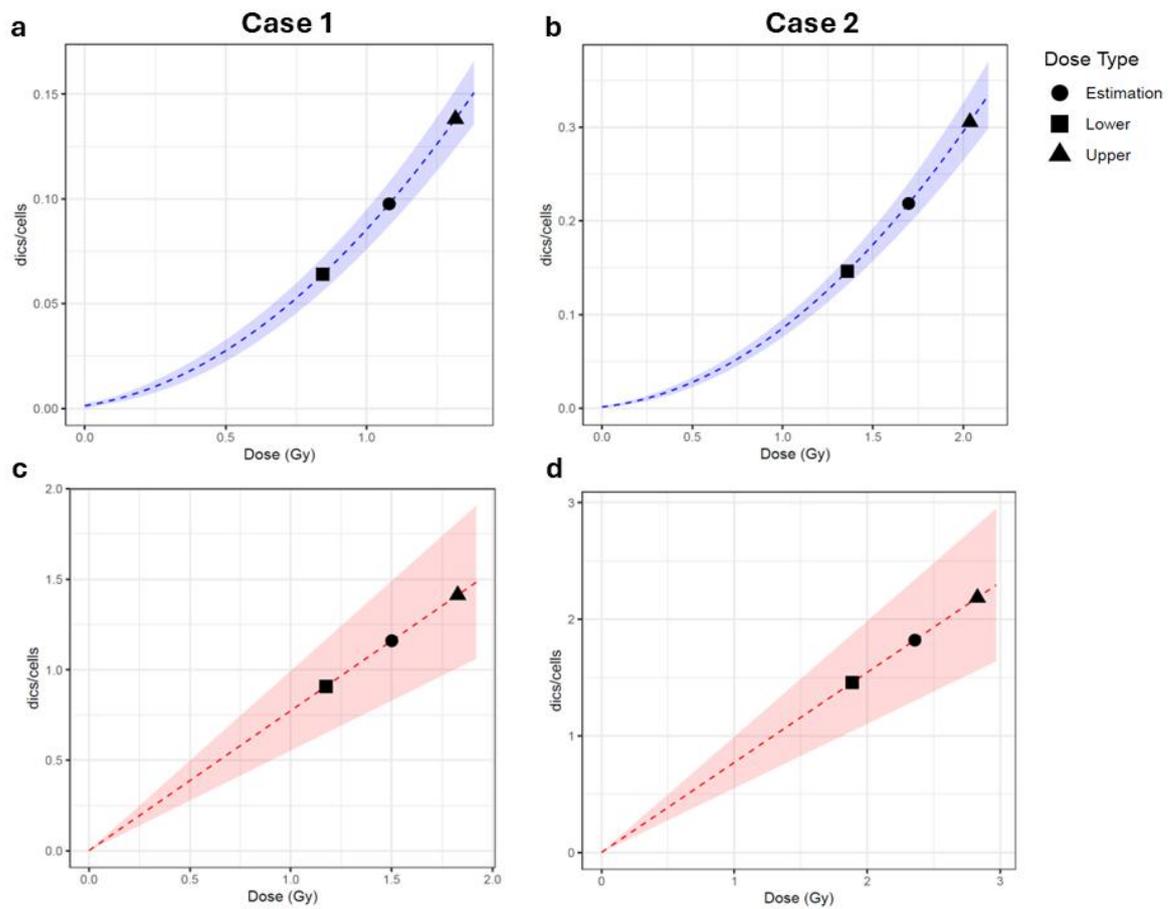

**Figure 2**. Gamma-ray (blue a,b) and Neutron curve (red c,d) plots with the estimated dose (circles) and the upper (triangles) and lower limits (squares). The confidence bands were calculated with the delta method.

**Figure 3**. 'Data input' box with three uploaded .rds files corresponding to a blind sample of the laboratories used in Bucher et al. 2025.

**Figure 4**. 'Z-Score data input' box with the reference value for the three samples analysed by each laboratory.

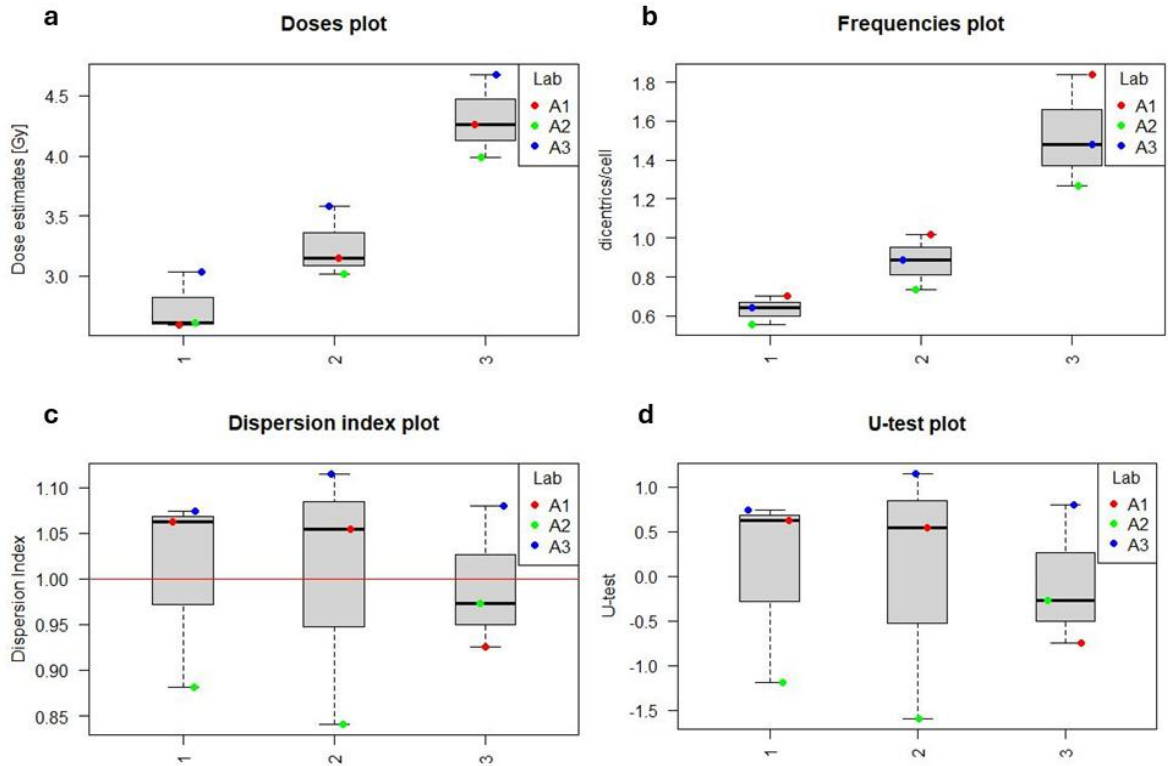

**Figure 5**. (a) Doses plot with upper and lower limits. (b) Frequencies plot with upper and lower limits. (c) Dispersion index plot. The red line represents the value 1 for the dispersion index ($\frac{\sigma^2}{y} = 1$). (d) U-test plot. A u-value > 1.96 indicates overdispersion, whereas a u-value < -1.96 suggests underdispersion (Rao and Chakravarti 1956; Savage 1970). All these graphical outputs work as a data summary and helps to visualize heterogeneity and any trend in dispersion of the results from all laboratories. Each laboratory (A1, A2, A3) is represented by a different color with a boxplot for each sample (1, 2, 3).

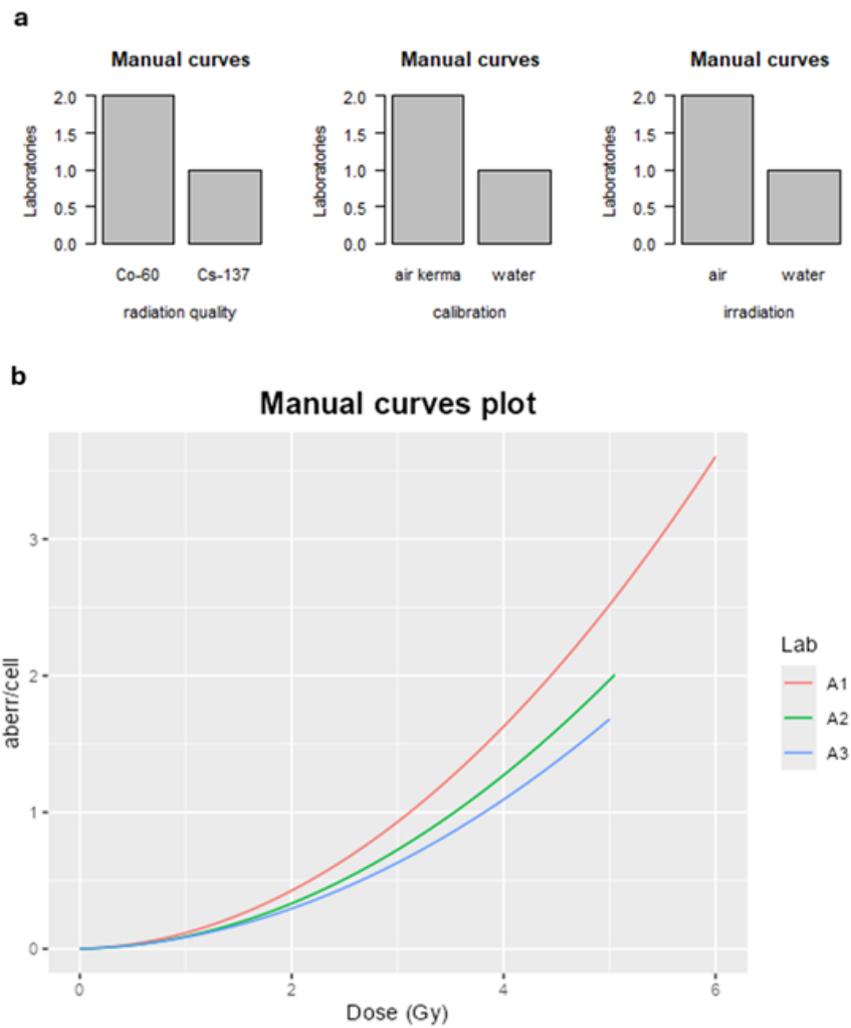

**Figure 6** (a) Bar plots for irradiation conditions from manual scored curves: Radiation quality, Calibration and Irradiation. (b) Manual scored curves by laboratory. These graphical outputs compare between calibration curves of participants and the differences in irradiation conditions for the establishment of the curves.

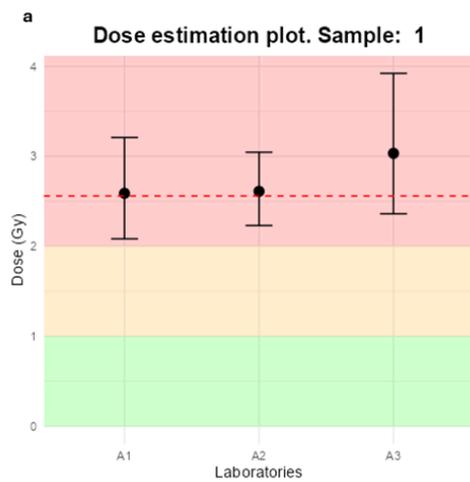
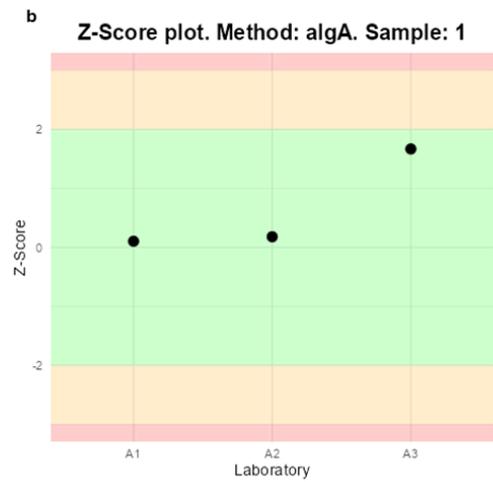
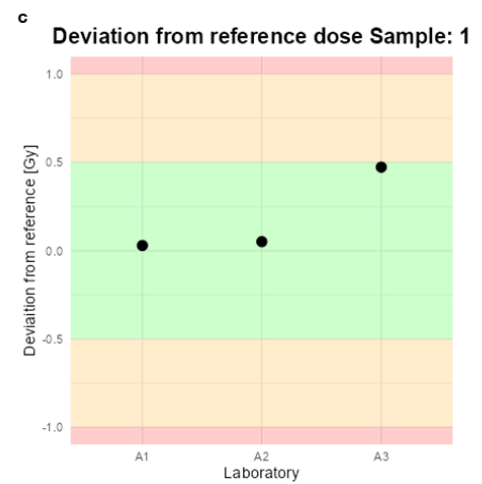
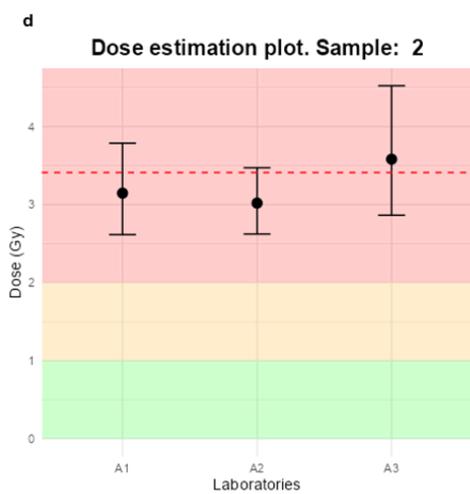
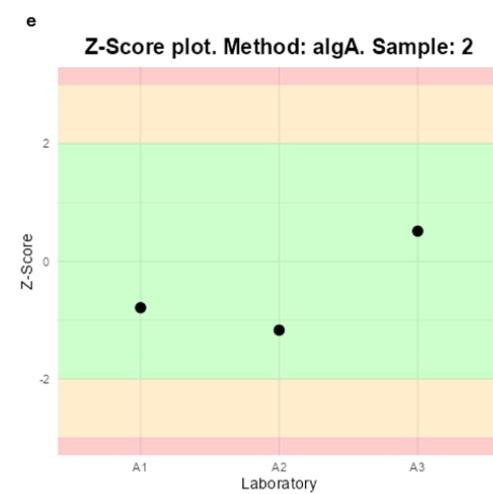
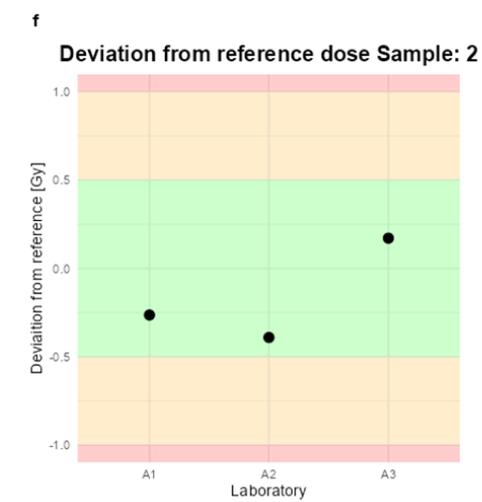
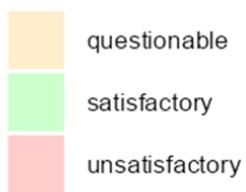
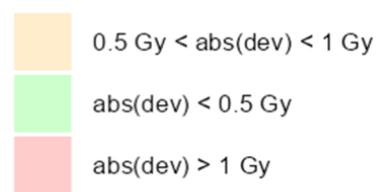

**Figure 7**. Dose estimation, Z-Score and Deviation from reference dose plots respectively for sample 1 (a)-(c) and for sample 2 (d)-(f). Separated plots for all samples to assess the performance of the laboratories. The plotting of Z-Scores with corresponding performance classification categories show which labs performed satisfactory, questionable or unsatisfactory. The plot for deviation from the reference dose shows the deviation in Gy for each laboratory and each sample and gives an impression on how close the dose estimates of each lab are to the reference dose.

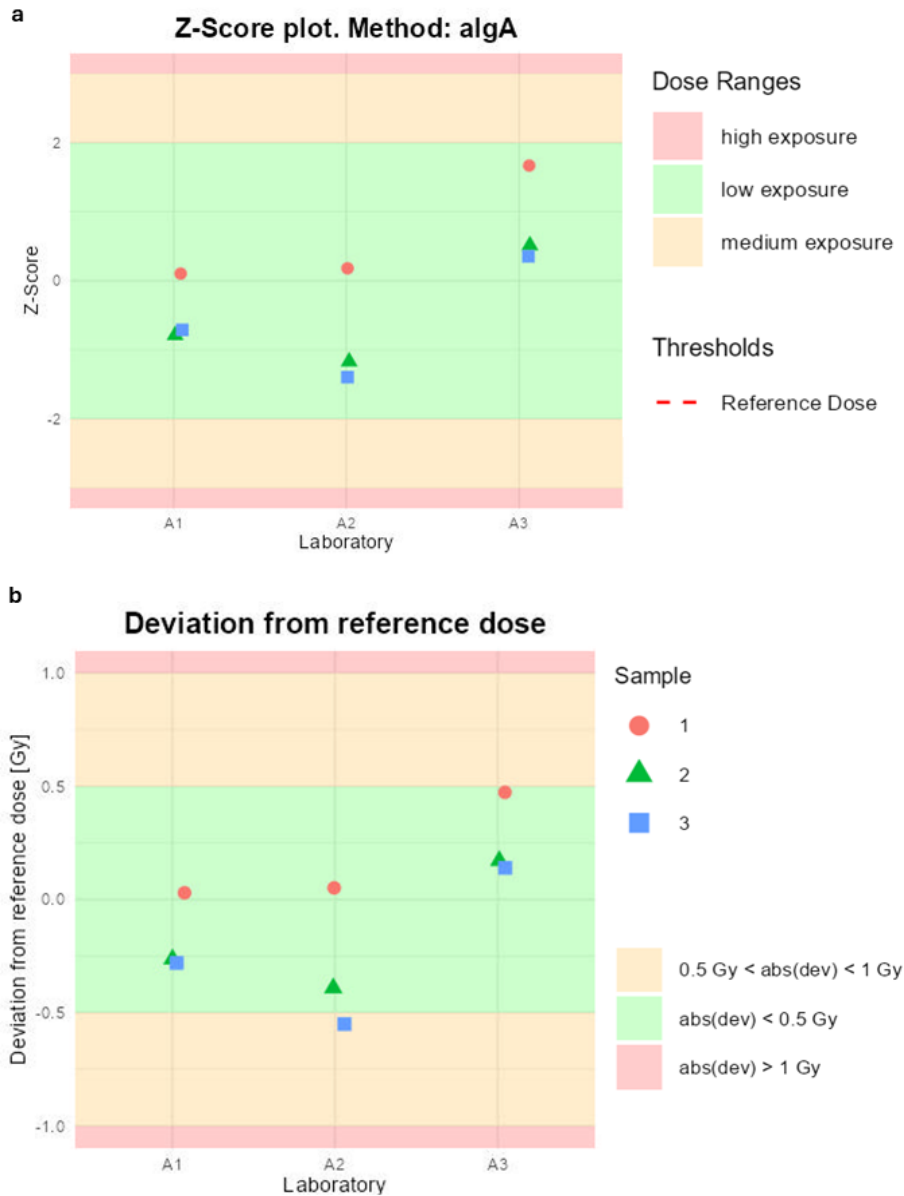

**Figure 8**. (a) Z-Score plot for all samples together with color status. (b) Deviation from reference plot for all samples together with color status. Z-Score plots and deviation from reference dose plots for all samples can show if there are systematic trends for a specific laboratory or all laboratories within an ILC.

## TABLES

**Table 1**. Output table for characteristic limits. In the column *Number cells case* a series of default numbers appear (20, 50, 100, 200, 500, 1000), but specific numbers can be selected too. This table also reports the decision threshold, the minimum number of significant dicentrics, the detection limit, and if a curve is provided, the minimum resolvable dose and the dose at detection limit (Gy).

| Number cells case | Decision threshold (y) | Min. dics significant (y+1) | Detection limit | Minimum resolvable dose (Gy) | Dose at detection limit (Gy) |
|---|---|---|---|---|---|
| *20* | 0 | 1 | 2.303 | 0.728 | 1.187 |
| *50* | 1 | 2 | 3.89 | 0.634 | 0.947 |
| *100* | 1 | 2 | 3.89 | 0.403 | 0.623 |
| *200* | 2 | 3 | 5.322 | 0.329 | 0.489 |
| *500* | 3 | 4 | 6.681 | 0.2 | 0.302 |
| *1000* | 7 | 8 | 11.771 | 0.2 | 0.274 |

Table 2. Output table for Z-Score using Dose as input type containing the following columns: Laboratory Name (*Lab*), Sample Name (*Sample*), Type of scoring (*Type*), Reference Value (*Reference*), Dose, Z-Score value (*Zscore*), Deviation from reference (*Deviation*) and Status.

| *Lab* | *Sample* | *Type* | *Reference* | *Dose* | *Zscore* | *Deviation* | *Status* |
|---|---|---|---|---|---|---|---|
| *A1* | 1 | manual | 2.56 | 2.5892 | 0.10293491 | 0.0292 | **Satisfactory** |
| *A2* | 1 | manual | 2.56 | 2.611 | 0.17978357 | 0.051 | **Satisfactory** |
| *A3* | 1 | manual | 2.56 | 3.0332 | 1.6681095 | 0.4732 | **Satisfactory** |
| *A1* | 2 | manual | 3.41 | 3.1461 | -0.78944264 | -0.2639 | **Satisfactory** |
| *A2* | 2 | manual | 3.41 | 3.0187 | -1.17055287 | -0.3913 | **Satisfactory** |
| *A3* | 2 | manual | 3.41 | 3.5812 | 0.51213558 | 0.1712 | **Satisfactory** |
| *A1* | 3 | manual | 4.54 | 4.2594 | -0.71172587 | -0.2806 | **Satisfactory** |
| *A2* | 3 | manual | 4.54 | 3.9892 | -1.39707273 | -0.5508 | **Satisfactory** |
| *A3* | 3 | manual | 4.54 | 4.6795 | 0.35383378 | 0.1395 | **Satisfactory** |

**SUPPLEMENTARY INFORMATION**

**S.I. Table 1.** Summary table generated in the Interlaboratory comparison module. This table contains (from left to right) the laboratories names, the module where the .rds file is obtained, the curve type, sample names (doses evaluated by each laboratory), the number of cells, the number of dicentrics, the dose estimation, lower and upper dose estimation limits, the mean (frequency), the standard error, the dispersion index and the u-test value.

| Lab | Module | Type | Sample | N | X | estimate | lower | upper | y | y.err | DI | u |
|---|---|---|---|---|---|---|---|---|---|---|---|---|
| A1 | dicentrics | manual | 1 | 200 | 140 | 2.5892 | 2.0834 | 3.2089 | 0.7 | 0.061 | 1.0625 | 0.6252 |
| A2 | dicentrics | manual | 1 | 200 | 111 | 2.611 | 2.2311 | 3.0456 | 0.555 | 0.0495 | 0.8818 | -1.184 |
| A3 | dicentrics | manual | 1 | 200 | 129 | 3.0332 | 2.3614 | 3.9212 | 0.645 | 0.0588 | 1.0735 | 0.7365 |
| A1 | dicentrics | manual | 2 | 200 | 204 | 3.1461 | 2.6149 | 3.7859 | 1.02 | 0.0733 | 1.0539 | 0.5389 |
| A2 | dicentrics | manual | 2 | 200 | 147 | 3.0187 | 2.6227 | 3.4713 | 0.735 | 0.0556 | 0.8406 | -1.5951 |
| A3 | dicentrics | manual | 2 | 200 | 177 | 3.5812 | 2.8631 | 4.5212 | 0.885 | 0.0702 | 1.1149 | 1.1496 |
| A1 | dicentrics | manual | 3 | 200 | 368 | 4.2594 | 3.8823 | 4.6889 | 1.84 | 0.0923 | 0.9255 | -0.7442 |
| A2 | dicentrics | manual | 3 | 200 | 253 | 3.9892 | 3.553 | 4.4873 | 1.265 | 0.0785 | 0.9731 | -0.2692 |
| A3 | dicentrics | manual | 3 | 200 | 296 | 4.6795 | 3.8797 | 5.7114 | 1.48 | 0.0894 | 1.0795 | 0.7939 |

**S.I. Table 2.** Curve's summary table generated in the Interlaboratory comparison module. This table contains (from left to right) the laboratories names, the module where the .rds file is obtained , the curve type, the radiation quality, the calibration of the source, the irradiation medium, the irradiation temperature, the dose rate (Gy/min), the curve origin, the values C, α and β of the curve and their respective standard errors and the maximum dose used for the curve.

| Lab | Module | Type | radiation quality | calibration | irradiation | temperature | dose rate | curve origin |
|---|---|---|---|---|---|---|---|---|
| *A1* | dicentrics | manual | Cs-137 | air kerma | air | 20 | 0.446 | own |
| *A2* | dicentrics | manual | Co-60 | air kerma | air | 37 | 0.27 | own |
| *A3* | dicentrics | manual | Co-60 | water | water | 37 | 1.07 | own |

| Lab | C | alpha | beta | C std.error | alpha std.error | beta std.error | max curve dose |
|---|---|---|---|---|---|---|---|
| *A1* | 0.0012 | 0.019 | 0.0969 | 0.0001 | 0.0041 | 0.0035 | 6 |
| *A2* | 0.0005 | 0.0142 | 0.0759 | 0.0005 | 0.0044 | 0.0027 | 5.05 |
| *A3* | 0.0013 | 0.021 | 0.063 | 0.0005 | 0.0052 | 0.004 | 5 |